\documentclass[aps,prl,preprint,showpacs,groupedaddress]{revtex4-1}

\usepackage[latin1]{inputenc}
\usepackage{graphicx}
\usepackage{dcolumn}
\usepackage{bm}
\usepackage{hyperref}

\begin{document}


\title{Design techniques for superposition of acoustic bandgaps using
fractal geometries}


\author{S. Casti\~neira-Ib\'a\~nez}
 \affiliation{Dpto. F\'isica Aplicada, Universidad Polit\'ecnica de Valencia.}

\author{V. Romero-Garc\'ia}
 \author{J.V. S\'anchez-P\'erez}
 \email{jusanc@fis.upv.es}
 \affiliation{Centro de Tecnolog\'ias F\'isicas: Ac\'ustica, Materiales y Astrof\'isica, Universidad Polit\'ecnica de Valencia.}

\author{L.M. Garcia-Raffi}
 \affiliation{Instituto Universitario de Matem\'atica Pura y Aplicada, Universidad Polit\'ecnica de Valencia.}


\date{\today}

\begin{abstract}
Research into properties of heterogeneous artificial materials,
consisting of arrangements of rigid scatterers embedded in a
medium with different elastic properties, has been intense
throughout last two decades. The capability to prevent the
transmission of waves in predetermined bands of frequencies
-called bandgaps- becomes one of the most interesting properties
of these systems, and leads to the possibility of designing
devices to control wave propagation. The underlying physical
mechanism is destructive Bragg interference. Here we show a
technique that enables the creation of a wide bandgap in these
materials, based on fractal geometries. We have focused our work
in the acoustic case where these materials are called
Phononic/Sonic Crystals (SC) but, the technique could be applied
any types of crystals and wave types in ranges of frequencies
where the physics of the process is linear.
\end{abstract}


\maketitle

Research into properties of heterogeneous artificial materials,
consisting of arrangements of rigid scatterers embedded in a
medium with different elastic properties, has been intense
throughout last two decades. The capability to prevent the
transmission of waves in predetermined bands of frequencies
-called bandgaps- becomes one of the most interesting properties
of these systems, and leads to the possibility of designing
devices to control wave propagation. The underlying physical
mechanism is destructive Bragg interference. Here we show a
technique that enables the creation of a wide bandgap in these
materials, based on fractal geometries. We have focused our work
in the acoustic case where these materials are called
Phononic/Sonic Crystals\cite{martinez95, Sanchez98} (SC) but, the
technique could be applied any types of crystals and wave types in
ranges of frequencies where the physics of the process is linear.

If we consider acoustically-hard cylinders (scatterers)
periodically embedded in air (host), then the difference between
velocities and densities in the scatterers and embedding medium
are very large. So the physical problem is reduced to that of
array scattering based on Bragg's law. With these conditions, the
position and the size of the bandgaps in the range of frequencies
depend on: a) the arrangement of the scatterers, according to the
Bragg law and b) the amount of matter formed by the scatterers,
quantified by the filling fraction (ff). For a given SC, an
increase in the bandgaps can be obtained only by an increasing the
ff. There are two main ways for varying the characteristics of
full SC bandgaps\cite{Khelif04, Sanchez02}. First, by varying the
acoustical properties of the scatterers\cite{Umnova06, Liu00,
Kuang04} or, secondly, by developing new arrangements of
scatterers with further crystalline symmetries. Quasi
crystals\cite{Zhang07} and Quasi Ordered Structures\cite{Romero06}
are examples of this second strategy. Also, other authors
\cite{Florescu09} have developed an optimized design technique
that could be applied to the case of elastic/acoustic waves, based
on the concept of hyperuniformity, to obtain large and complete
bandgaps with amorphous photonic materials. Here, we propose a new
way to obtain large bandgaps based on the redistribution of the
elements of the SC based on fractal geometries\cite{Mandelbrot83}.
We have chosen these geometries because they can be modelled
mathematically and they can be used as design tools. Recently
fractals have been under study for a wide range of practical
applications, from biological or medical\cite{Iannaccone96} to
economics\cite{Williams95}. In fact, fractals have been used for
SCs\cite{Frezza04, Liang03, Fu02, Zheng03} but only to design of
the shapes of the scatterers\cite{Norris08}.

\begin{figure}[hbt]
\includegraphics[width=80mm,height=100mm,angle=0]{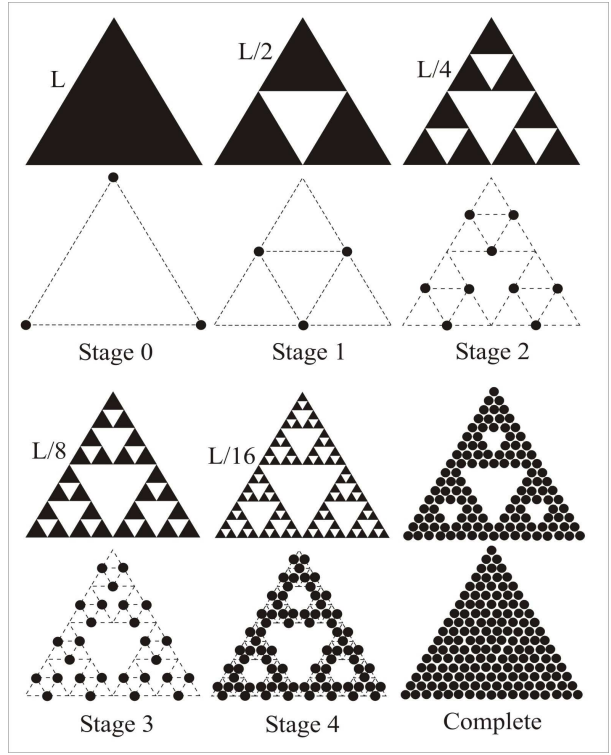}%
\caption{\label{fig:Figure1}Quasi fractal arrangement of
scatterers: Five stages of cylinder arrays based on the
Sierpinski´s triangle geometry and the resulting complete
structure.}
\end{figure}

As the first step we have designed an arrangement of scatterers
inside an equilateral triangle of side L, based on a 2D fractal
called Sierpinsky triangle (see figure \ref{fig:Figure1}). We have
chosen a 2D symmetry triangular pattern that presents the highest
bandgap as a consequence of its degree of
hyperuniformity\cite{Florescu09}. In this figure, we represent a
transversal section (in the XY plane) of our arrangement assuming
infinitely long cylinders with radius r parallel to the z axis. We
have called it Quasi Fractal Structure (QFS) because, although the
fractal construction follows an infinite iterative
process\cite{Mandelbrot83}, we only show here the first five
iterations to take account of the space restrictions given by both
L and r.  Figure \ref{fig:Figure1} shows also that a cylinder is
located at every vertex of the empty triangles in each stage,
except stage zero where we have located the scatterers at the
vertex of the existing triangle. Also, figure \ref{fig:Figure1}
shows the sum of the different stages of our fractal arrangement
(complete figure). At first glance one might think that it is a
classical triangular crystalline array with some vacancies in its
structure. However, the underlying symmetry follows a fractal
pattern. Thus, we can consider the complete figure as a sum of
independent triangular arrays with different lattice constants
($L$, $L/2$, $L/4$, $L/8$ and $L/16$), with every stage located
iteratively within the previous one. This provides a compact small
device and the obtained resultant full bandgap results from the
sum of the Bragg peak corresponding to every array. This idea is
consistent with the nature of fractal geometries based as they are
on the repetition of identical motifs at differing size
scales\cite{Mandelbrot83}.

Another argument to explain the existence of a large full bandgaps
is related to the relationship among the different lattice
constants. Here they are proportional to $1/2^M$, being $M$ the
order number of the stage. This produces a repetition of many
Bragg peaks at different stages and a reinforcement of the
bandgap. It is possible to find an expression to obtain the number
of repeated Bragg peaks at different stages. The following
functions $S_{\alpha}(n, M)$, $\alpha=0^{\circ}$, 30$^{\circ}$
give the value of the frequency for which the n-th Bragg peak
appears at the different stages $M$ $(M = 0, 1, 2, 3, 4)$, as a
function of $L$ and along the two high-symmetry directions of the
triangular array (0$^{\circ}$, 30$^{\circ}$)

\begin{eqnarray}
S_{0^{\circ}} (n,M)=C_{0^{\circ}}(n+1)2^M;\nonumber\\
S_{30^{\circ}}(n,M)= C_{30^{\circ}}(n+2)2^{M-1},\label{eq:1}
\end{eqnarray}

where $C_{0^{\circ}}=\sqrt(3)/3$ and $C_{30^{\circ}}=2/3$ due to
the Bragg law. Based on expressions (\ref{eq:1}), it is
straightforward to find the relationship of appearance of a
predetermined Bragg peak for any two different stages

\begin{eqnarray}
(0^{\circ})\;\;\;\; n=(n'+1)2^V-1;\nonumber\\
(30^{\circ})\;\;\;\; n=(n'+2)2^V-2, \label{eq:2}
\end{eqnarray}

where $V$ is the difference between the couple of stages we want
to compare $(V=1, 2, 3, 4)$. Equations (\ref{eq:2}) show the
relationship between the n-position of appearance of a Bragg peak
in the stage $M$ as a function of the $n'$ position of appearance
of the same peak in another stage $M'$, such that $V=M'-M$. Note
the large number of times certain Bragg peaks are repeated on
different stages according with (\ref{eq:2}), producing an
reinforcement and, as a consequence, an enhancement of the full
bandgap.

\begin{figure}[hbt]
\includegraphics[width=140mm,height=60mm,angle=0]{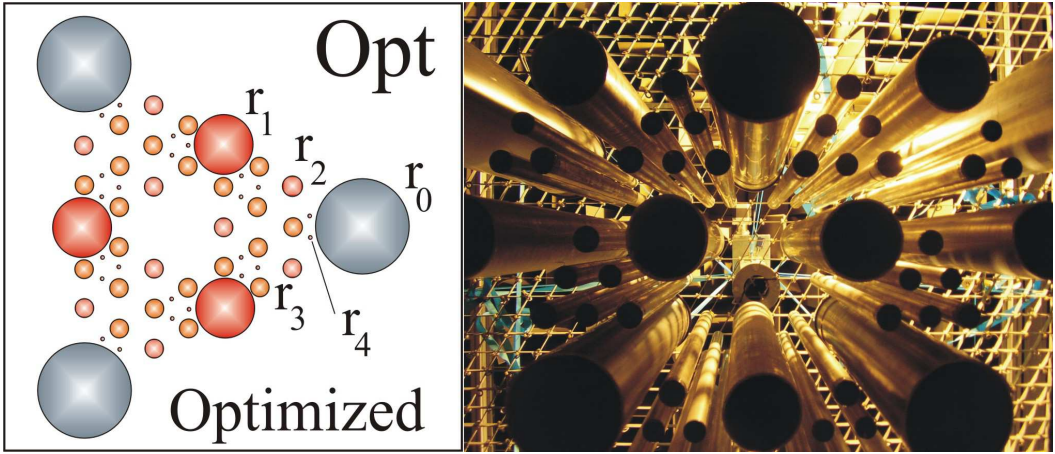}%
\caption{\label{fig:Figure2}(Left) Optimized arrangement of
scatterers based on Sierpinski triangle with different
relationships among the radii of the cylinders (QFS$_{Opt)}$;
(Right) Photograph taken from beneath the commercial arrangement
(QFS$_{Exp}$) used to validate theoretical results. Part of the
supporting frame can be seen also.}
\end{figure}

The second, and much more important, step of our design technique
consists in varying the diameter of each set of cylinders for each
stage independently. Thus, the scatterers are distributed in a
more efficient way, increasing the sizes in the large stages and
reducing them in the others, thereby providing each stage with the
adequate value of ff for the appearance of their Bragg´s peaks. As
a consequence, a further increase of the full bandgap occurs. In
figure \ref{fig:Figure2} we show a proposed QFS built with an
optimized relationship between the radii of the cylinders
belonging to the different stages $M$ $(M=0, 1, 2, 3, 4)$. For the
optimization process we have used genetic algorithm already
adapted to the acoustic case\cite{Romero06} (QFS$_{Opt}$)
($r_0/L\approx0.14$; $r_1/L\approx0.09$; $r_2/L\approx0.03$;
$r_3/L\approx0.032$; $r_4/L\approx0.02$). Note that it has been
necessary to remove some cylinders of the starting complete array
shown in figure \ref{fig:Figure1} in order to place the biggest
cylinders (large radii) of first stages. Of course, other
relationships among the radii of the cylinders could be
appropriate other applications.

\begin{figure}[hbt]
\includegraphics[width=130mm,height=100mm,angle=0]{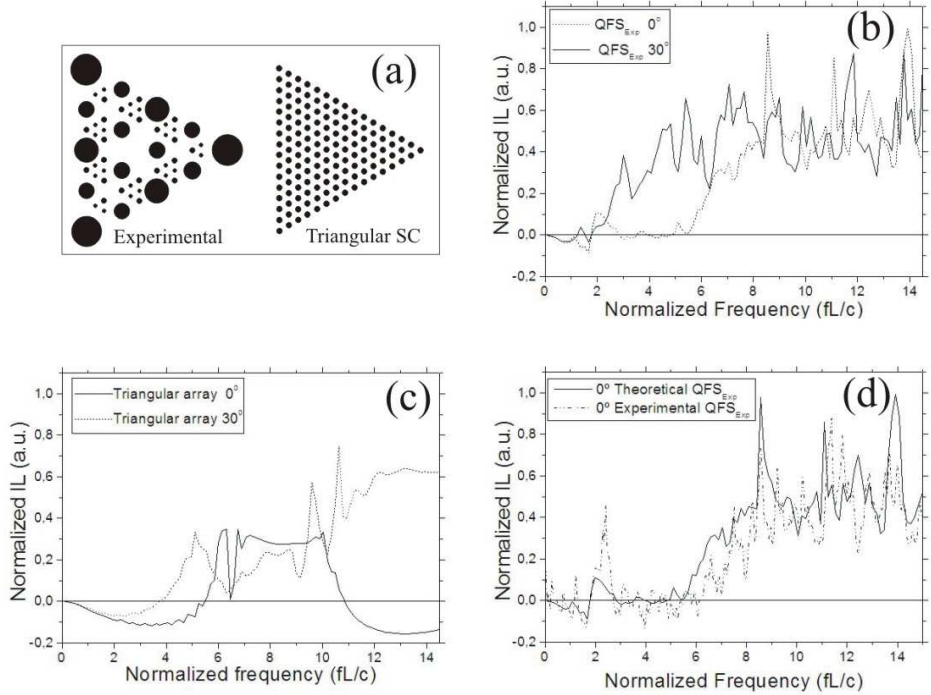}%
\caption{\label{fig:Figure3}Non-transmission properties of the
designed devices: (a) Experimental Quasi Fractal Structure
(QFS$_{Exp}$) and the triangular SC used. The relationship among
the radii of the cylinders for the first case is
($r_0/L\approx0.094$; $r_1/L\approx0.078$; $r_2/L\approx0.054$;
$r_3/L\approx0.029$; $r_4/L\approx0.017$), being
(ff$_{Exp}$=33\%); (b) Theoretical normalized IL spectra along the
two high-symmetry directions for QFS$_{Exp}$. (c) The same for the
triangular array (ff$_{SC}$=36\%); (d) Both Theoretical and
Experimental QFS$_{Exp}$ attenuation spectra for $\Gamma$X
direction.}
\end{figure}

To quantify the size of the bandgap of this device we have used
the Attenuation Area parameter\cite{Romero06} (AA) in the analyzed
range of frequencies and, at the moment, only along the $\Gamma$X
direction (0$^{\circ}$ of incidence on the sample). Comparing the
AA value for the QFS$_{Opt}$ designed with the corresponding to a
classical SC with triangular array constructing with the same
external size and shape and with cylinder radius $r/L\approx0.02$,
we obtain interesting results: AA parameter grows, in QFS$_{Opt}$
case (AA$_{Opt}$=179.88 normalized units) more than 400\% compared
with the classical triangular lattice (AA$_{SC}$=43.94n.u.). But
keep in mind that QFS$_{Opt}$ has been designed under the premise
of maintaining the same ff as SC by means of genetic algorithms
(ff$_{Opt}$=ff$_{SC}$=36\%). With these data, we can break the
rule about the relationship between ff and the size of the
bandgaps: we have obtained a high increase in the size of the
bandgap without increasing the ff of the device in respect of the
original triangular array device. These results have been
calculated in the normalized range of frequencies 0-15 shown in
figure \ref{fig:Figure3}. Moreover, due to the nature of our
technique, the crystal wave properties of our device remain intact
for each stage as it is a sum of triangular arrays. This means
that both the bandgap obtained along the other high-symmetry
direction $\Gamma$J (30$^{\circ}$) and the full bandgap also grow
(200\% in the case of the full bandgap).

To illustrate the above statement experimentally we have
constructed a new device similar to QFS$_{Opt}$ but with
commercially available hollow cylinders, QFS$_{Exp}$ (figure
\ref{fig:Figure3}a). In figures
\ref{fig:Figure3}b-\ref{fig:Figure3}c one can compare the
theoretical normalized insertion loss spectra (IL), along the two
high-symmetry directions $\Gamma$X and $\Gamma$J
(0$^{\circ}$-30$^{\circ}$), for both QFS$_{Exp}$ and the SC
defined above. We have used Multiple Scattering
Theory\cite{Chen01, Garcia00} to obtain these spectra, which have
been calculated at a distance $d=1/L$ from the edge of the
samples. Also, in figure \ref{fig:Figure3}d we show the good
agreement between the theoretical and experimental results for
0$^{\circ}$ incidence.

In summary, in this work we have shown that an optimised
fractal-based design technique enables a large increase of the
scattering bandgaps for sonic crystal arrays if rigid scatterers.
There are two steps. The first consists of the use of fractal
patterns to arrange the scatterers. The resulting device is the
sum of several independent crystalline arrays. The second step
consists of optimising the nested arrays by varying the ff of each
fractal stage independently. As a result, we have obtained
efficient and compact devices. The sum of the Bragg peaks
belonging to the different scale arrays (stages), the
reinforcement process due to the existence of different lattice
constants and the redistribution of cylinders among the different
stages are behind this enhancement.

\begin{acknowledgments}
This work was supported by MCI (Spanish Government) and FEDER
funds, under Grant Nos. MAT2009-09438 and MTM2009-14483-C02-02.
The authors would like to thank Prof. K. Attenborough and Dr. E.A.
Sánchez-Pérez for their suggestions and for the revision of the
manuscript.
\end{acknowledgments}

%

\end{document}